\documentclass{revtex4-1}

\usepackage{graphicx}
\usepackage{amssymb}
\usepackage{amsmath}
\usepackage{color}
\usepackage{ wasysym }
\usepackage{hyperref}
\usepackage{tabu}

\def\barray{\begin{array}}
	\def\earray{\end{array}}
\def\be{\begin{equation}}
\def\ee{\end{equation}}
\def\ben{\begin{equation} \nonumber}
\def\een{\end{equation}}
\def\ban{\begin{eqnarray*}}
	\def\ean{\end{eqnarray*}}
\def\ba{\begin{eqnarray}}
\def\ea{\end{eqnarray}}

\def\({\left(}
\def\){\right)}

\def\physrep{Phys. Rep.}

\newcommand{\bse}{\begin{subequations}}
\newcommand{\ese}{\end{subequations}}
\def\barray{\begin{array}}
\def\earray{\end{array}}
\def\be{\begin{equation}}
\def\ee{\end{equation}}
\def\ben{\begin{equation} \nonumber}
\def\een{\end{equation}}
\def\ban{\begin{eqnarray*}}
\def\ean{\end{eqnarray*}}
\def\bea{\begin{eqnarray}}
\def\eea{\end{eqnarray}}

\graphicspath{{./fig/}}

\RequirePackage{color}

\begin{document}

\title{Gaugessence: a dark energy model with early time radiation-like equation of state}

\author{Ahmad Mehrabi}
\email{Mehrabi@basu.ac.ir}
\affiliation{Department of Physics, Bu-Ali Sina University, Hamedan, 65178, 016016, Iran}

\author{Azade Maleknejad}
\email{azade@ipm.ir}
\affiliation{School of Physics, Institute for Research in Fundamental Sciences (IPM),\\ P. Code. 19538-33511, Tehran, Iran}

\author{Vahid Kamali}
\email{vkamali@basu.ac.ir}
\affiliation{Department of Physics, Bu-Ali Sina University, Hamedan, 65178, 016016, Iran}
\affiliation{School of Physics, Institute for Research in Fundamental Sciences (IPM),\\ P. Code. 19538-33511, Tehran, Iran}

\begin{abstract}
In this work, 
we study a new quintessence model associated with non-Abelian gauge fields, minimally coupled to Einstein gravity. This gauge theory has been recently introduced and studied as an inflationary model, called gauge-flation. Here, however, we are interested in the late time cosmology of the model in the presence of matter and radiation to explain the present time accelerating Universe. During the radiation and matter eras, the gauge field tracks radiation and basically acts like a dark radiation sector. As we approach lower redshifts, the dark component takes the form of a dark energy source which eventually becomes the dominate part of the energy budget of the Universe. Due to the tracking feature of our model, solutions with different initial values are attracted to a common trajectory. The existence of early dark radiation is a robust prediction of our model which contributes to the effective number of relativistic species, $N_{\rm eff}$ and has its own interesting observational features.
\end{abstract}

\maketitle

\section{Introduction}

The late time cosmic acceleration first confirmed by supernovae Ia (SnIa) surveys \cite{Riess:1998cb, Perlmutter:1998np, Riess:1998dv}. Moreover, other observations including large scale structure (LSS) \cite{Tegmark:2003ud, Tegmark:2006az}, cosmic
microwave background (CMB) \cite{Spergel:2003cb, Ade:2015xua, Ade:2015rim} and baryon acoustic oscillations (BAO) \cite{Percival:2007yw, Aubourg:2014yra} measurements indicate that about 70\% of the energy density of the Universe today consists of an unknown component called dark energy (DE). The quest for the nature of DE is one of the most exciting challenges of the modern cosmology. Various observational projects are planned or envisioned to shed light on the physics underlying the observed cosmic acceleration and in the near future, we expect a wealth of new high quality data on that direction \cite{Huterer:2013xky, Hojjati:2015qwa}.

The cosmological constant with equation of state $w\equiv\frac{P_{DE}}{\rho_{DE}}=-1$ seems the most simple explanation for the late time accelerated expansion. However, if the cosmological constant is responsible for the current epoch of acceleration, its value ($\Lambda\sim 10^{-120}M_{\rm pl}^4$) is many orders of magnitude smaller than the theoretically expected value \cite{Weinberg:1988cp, Nobbenhuis:2006yf} so the  value of cosmological constant should extremely tuned at early time. Anyway such problem arises in any DE models. It could be an unknown energy component in the universe, DE \cite{Sahni:1999gb, Peebles:2002gy, Carroll:2000fy, Tsujikawa:2010sc, Copeland:2006wr}, or the modification of gravity as described by Einstein's general relativity, modified gravity  (MG) \cite{DeFelice:2010aj, Tsujikawa:2010zza, Clifton:2011jh, Amendola:2012ys}. Those models lead to a similar late time expansion history as the cosmological constant with a much richer phenomenology \cite{Amendola:2012ys, 2012PhR...513....1C, Song:2015oza}.

Many of the dark energy and modified gravity models can be described by adding a scalar degree of freedom to Einstein gravity \cite{Gubitosi:2012hu}.
From the former viewpoint, one possibility is a canonical scalar field which varies slowly along its potential, known as the standard quintessence model \cite{Copeland:2006wr, Tsujikawa:2013fta, Zlatev:1998tr} with the possibility of link to the models of particle physics. On the other hand, dynamical gauge fields abound in models of particle physics, string theory, GUTs and etc. It is therefore natural to ask whether gauge fields can be involved in the new physics beyond vanilla $\Lambda$CDM, responsible for the present dark energy era. Inspired by that idea, one may explore the possibility of using non-Abelian gauge fields as a dark energy candidate.

Over the past decade, there have been interesting discussions in the literature of using non-Abelian gauge fields in inflationary models \cite{Maleknejad:2011jw, Maleknejad:2011sq, SheikhJabbari:2012qf,Bielefeld:2015daa}, models of dark energy \cite{Gal'tsov:2010dd, Rinaldi:2015iza,Elizalde:2012yk,Bamba:2008xa,Setare:2013dra,Setare:2013kja} and dark matter \cite{Buen-Abad:2015ova, Gross:2015cwa, Zhang:2009dd,Foot:2014uba,Foot:2014osa,Foot:2016wvj}. For a detailed review of literature on the gauge fields in inflation, see \cite{Maleknejad:2012fw}. The interest in non-Abelian gauge fields is due to the fact that for a generic gauge invariant field theory, there is an isotropic and homogeneous field configuration for the SU(2) gauge field \cite{Gal'tsov:2010dd,Maleknejad:2011jw}.
In particular, choosing the temporal gauge $A^a_0=0$, we can identify the SU(2) gauge index with the spatial rotation SO(3) index in the spatial elements of the field. The possibility of that identification then leads to an isotropic and homogeneous background solution of the form $A^a_i=\phi(t)\delta^a_i$. The next generic feature of gauge field models is that if the theory consists of \textit{only} the Yang-Mills term, then the gauge field will track the radiation and damps like $a^{-4}$ by the expansion of the Universe. In order to prevent that, one should break the conformal symmetry by coupling the gauge field to the other fields, e.g. Higgs boson \cite{Rinaldi:2015iza} and axion field \cite{Adshead:2012kp, Martinec:2012bv}, or adding new gauge invariant terms to the gauge field theory \cite{Maleknejad:2011jw}.

Among the interesting predictions of models including gauge fields is the existence of a weakly interacting dark radiation at high redshifts. Presence of additional relativistic degrees of freedom leads to important modifications comparing to the $\Lambda$CDM, e.g. changing the Hubble parameter and observable signals in the CMB \cite{Baumann:2015rya, Ackerman:mha}. The observational measurements ($N_{\rm eff}=3.15\pm0.23$) have been used to constrain the effective value of extra relativist degrees of freedom, with the current bound $\Delta N_{\rm eff}=0.1\pm0.23$ \cite{Ade:2015rim}. On the theoretical point of view, most of inflationary and dark matter models motivated by string theory and particle physics beyond SM predict a form of dark radiation \cite{Lesgourgues:2015wza, Ackerman:mha, Buen-Abad:2015ova}. Future CMB polarization experiments will improve our constraints on these new physics by one or two order of magnitude \cite{Wu:2014hta}. This significant improvement in sensitivity of the future observational measurement, further fueled the interest in dark radiation.

In the present paper, we study a new quintessence model associated with non-Abelian gauge fields, gaugessence. Our gauge field theory has been first presented and investigated as an inflationary model (gauge-flation) in which inflation is driven by non-Abelian gauge fields \cite{Maleknejad:2011jw, Maleknejad:2011sq}. Here, we are interested in the late time cosmology of the model in the presence of matter and to explore if this model can explain the present time accelerating Universe. Our gauge field theory consists of the Yang-Mills which acts like a dark radiation term and $\kappa(F\tilde F)^2$ with the equation of a state $w=-1$. Our model is parametrized by only two parameters, the gauge field coupling $g$ and the dimensionfull prefactor $\kappa$. We are working on weakly coupled theories, thus, confinement in the dark sector is irrelevant.

This paper is organized as follows. In section \ref{setup}, we introduce the setup of the gaugessence model. We study the cosmological evolution of the model analytically in section \ref{analytical}. Section \ref{latetime} studies the late time cosmology of the model and presents the result of numerical analysis of the DE trajectories. In section \ref{data-model}, we confront the model with the observational data and show the $1\sigma$ and $2\sigma$ confidence regions  of the model parameters. Section \ref{pert} is devoted to the qualitative study of the cosmic perturbation in our model.Finally we conclude in section \ref{conclusion}. Throughout this paper, we set the reduced Planck mass $M^2_{pl}=(8\pi G)^{-1}$, equal to one.

\section{Gaugessence model}\label{setup}

The gaugessence model consists of radiation, baryonic and cold dark matter (CDM), as well as a dark non-Abelian gauge field sector. The hidden gauge field sector is minimally coupled to Einstein gravity and interacts gravitationally with the visible matter. This model is specified by the following action
\begin{eqnarray}\label{s}
S&=&\int\sqrt{-g}\bigg(\!-\frac{1}{2}R\\ \nonumber &-& \frac{1}{4}\big(F_{\mu\nu}^{a}F_{a}^{\mu\nu}+\frac{\kappa}{24}(F^{a}_{\mu\nu}\tilde{F}^{a\lambda\sigma})^2\big)\bigg)d^4x+S_m(\chi_i,g_{\mu\nu}),
\end{eqnarray}
where $R$ is the Ricci scalar, $\tilde{F}^{a\mu\nu}=\frac12\epsilon^{\mu\nu\lambda\sigma}F^a_{\lambda\sigma}$ and $S_m$ is the matter action, including cold dark matter, baryons, photons and neutrinos. The gaugessence action consists of the Yang-Mills and the $(F\tilde F)^2$ term and has two parameters $\kappa$ and $g$.
The gauge field strength $F^a_{\mu\nu}$ is
\be
F^a_{\mu\nu}=\partial_{\mu}A^a_{\nu}-\partial_{\nu}A^a_{\mu}-g\epsilon^a_{bc}A_{\mu}^bA_{\nu}^c,
\ee
 where $\epsilon_{abc}$ is the totally antisymmetric tensor, $g$ is the gauge coupling and $A_{\mu}^a$ is a 4-dimensional SU(2) gauge field. Note that here $\mu,\nu,...$ and $a,b,...$ are respectively used for the indices of the space-time and the gauge algebra.
The gauge field theory $\mathcal{L}_{G}$ has been originally introduced and investigated as an inflationary model in \cite{Maleknejad:2011jw, Maleknejad:2011sq}, \textit{gauge-flation} model. Here, on the other hand, we are interested in its late time evolution and whether it gives rise to the late time cosmic acceleration as a quintessence model. Note that the specific term $(F\tilde{F})^2$ is chosen to generate the  acceleration of Universe and we will see that the contribution of this term to the energy-momentum tensor helps us to have $\rho+3P<0$. Obviously $\kappa$ in Eq.(\ref{s}) has negative mass dimension so the theory is non-renormalizable but like most theories, one can consider this theory as an effective theory of quantum gauge field. (for more discussion see \cite{Maleknejad:2011sq, Maleknejad:2012fw})

For later convenience, it is useful to decompose the energy and the pressure density of the gauge field theory into dark radiation (DR) and dark energy (DE) parts
\begin{align}\label{rho-p}
\rho_{G}=\rho_{DR}+\rho_{DE}\quad \textmd{and}\quad P_{G}=\frac13\rho_{DR}-\rho_{DE},
\end{align}
where $\rho_{DE}$ and $\rho_{DR}$ are the contributions of $(F\tilde F)^2$ and Yang-Mills terms respectively. In fact, DR is the contribution of the Yang-Mills, while the $(F\tilde F)^2$ is effectively the dark energy (DE) sector.
Thus, the hidden gauge field has a time dependent equation of state 
\be\label{w}
w(t)=\frac{\frac13\rho_{DR}-\rho_{DE}}{\rho_{DR}+\rho_{DE}}.
\ee
Thus in the regime that $\rho_{DE}\!\gg\!\rho_{DR}$, the gauge sector is effectively a dark energy element. Note that EoS is bounded in the range $-1< w(t)<\frac13$ and it never crosses the phantom line.

\subsection{Background evolution}\label{BG}
We now turn to investigate the background trajectories of the gaugessence model.
Considering the flat, isotropic and homogeneous FLRW metric
\be
\label{FLRW}
ds^2=-dt^2+a(t)^2\delta_{ij}dx^{i}dx^{j},
\ee
 we have radiation and matter energy densities in terms of their energy density at the present time as $\rho_{\rm r}(t)=\frac{\rho^0_{\rm r}}{a^4}$ and $\rho_{\rm m}(t)=\frac{\rho^0_{\rm m}}{a^3}$. To understand the dynamics of the gauge field sector, however, we need to solve the field equation for the isotropic and homogeneous space-time.
The non-Abelian gauge field has the following isotropic and homogeneous configuration \cite{Maleknejad:2011jw, Maleknejad:2011sq}
\be\label{ansatz}
A^a_0=0 \quad \textmd{and} \quad A^a_i=\phi(t)\delta_i^a,
\ee
where $i,j=1,2,3$ are labels of spatial directions. For a more comprehensive detailed discussion on the above ansutz see \cite{Maleknejad:2012fw}.
Using field configuration \ref{ansatz} and FLRW metric \ref{FLRW},
one can find the isotropic and homogeneous strength tensor
\begin{align}\label{Lagrangian}
F^a_{0i}=\dot{\phi}\delta_i^a \quad \textmd{and} \quad F^a_{ij}=-g\phi^2\epsilon_{ij}^a.
\end{align}
Moreover, the reduced action of the dark gauge field theory is
\begin{align}
&\mathcal{L}_{G}=\frac{3}{2}(\frac{\dot{\phi}^2}{a^2}-\frac{g^2\phi^4}{a^4}+\kappa g^2\frac{\dot{\phi}^2\phi^4}{a^6}),
\end{align}
while the field equation of the gauge field ($D_{\mu}(\frac{\delta \mathcal{L}}{\delta F_{\mu\nu}})=0$) can be determined as
\be\label{EOM1}
(1+\kappa g^2\frac{\phi^4}{a^4})\frac{\ddot{\phi}}{a}+(1+\kappa\frac{\dot{\phi}^2}{a^2})\frac{2g^2\phi^3}{a^3}+
(1-3\kappa g^2\frac{\phi^4}{a^4})\frac{H\dot{\phi}}{a}=0.
\ee
Recalling \eqref{rho-p}, the energy density of the gauge field can be read as $\rho_{G}=\rho_{DR}+\rho_{DE}$, in which 
\bea
\rho_{\rm DE}=\frac32\kappa\frac{g^2\dot{\phi}^2\phi^4}{a^6}\quad \textmd{and}\quad \rho_{\rm DR}=\frac{3}{2}(\frac{\dot{\phi}^2}{a^2}+\frac{g^2\phi^4}{a^4}).
\eea
In the presence of matter and radiation in a spatially flat universe, the Friedmann equations are 
\bse\label{Friedmann}
\begin{align}
3\big(\frac{\dot{a}}{a}\big)^2&=(\rho_{\rm m}+\rho_{\rm r}+\rho_{DR}+\rho_{DE}),\\
3\big(\frac{\ddot{a}}{a}\big)&=-(\frac12\rho_{\rm m}+\rho_{\rm r}+\rho_{DR}-\rho_{DE}).
\end{align}
\ese
Working out the field equations, now we turn to the further study of the gaugessence model during the matter and dark energy epochs.

\section{Cosmological evolution, analytical treatment}\label{analytical}

Since we are not able to measure distance directly, but redshift, it is more convenient to rewrite the fields in terms of redshift $z$ ($a=\frac{1}{1+z}$). 
Here, for further simplification, we define the following dimensionless parameters
\be\label{re-def}
\tilde g=\frac{g}{H_0},\quad  \tilde{\kappa}=g^2\kappa,\quad \textmd{and}\quad \tilde t=H_0t,
\ee
where $H_0$ is the present value of the Hubble constant \cite{Ade:2015rim}
\be
H_0=(67.8\pm0.9)\textmd{km}\textmd{s}^{-1}\textmd{Mpc}^{-1},
\ee
which is $H_0\simeq10^{-60}M_{pl}$.
Moreover, the energy density parameter of the gauge field, $\Omega_{G}(z)$, is defined as
\be\label{eq:omega-g}
\Omega_{G}(z)\equiv\frac{\rho_G}{\rho_{c}}=\Omega_{DR}(z)+\Omega_{DE}(z) \quad \textmd{where}\quad \rho_c=3H^2_0,
\ee
where $\Omega_{DR}(z)$ and $\Omega_{DE}(z)$ are respectively the contributions of the dark radiation and dark energy sectors
 \begin{eqnarray}\label{omegaG}
 \Omega_{DR}(z)&=&\frac12\bigg(\phi'^2(1+z)^2+
 \tilde{g}^2\phi^4(1+z)^4\bigg)\quad \\ \nonumber \quad \Omega_{DE}(z)&=&\frac12\bigg(\tilde\kappa\phi^4\phi'^2(1+z)^6\bigg),
 \end{eqnarray}
 where ``$'$'' means a derivative with respect to $\tilde t$.
Using \eqref{re-def} and \eqref{omegaG}, we can rewrite the Friedmann equations \eqref{Friedmann}, as
\bse\label{Frid}
\begin{align}
&(\frac{z'}{1+z})^2=\Omega_{\rm DE}(z)+\Omega_{\rm DR}(z)+\Omega_{\rm m}^0(1+z)^3+\Omega_{\rm r}^0(1+z)^4, \label{frid1}\\
&\frac{z''}{1+z}=3(\frac{z'}{1+z})^2-\frac12\Omega_{\rm m}^0(1+z)^3-2\Omega_{\rm DE}(z),\label{frid2}
\end{align}
\ese
as well as the field equation of $\phi$ \eqref{EOM1} which is
\bea\label{Phi}
&\big(&1+\tilde\kappa\phi^4(1+z)^4\big)\phi''+ 2\tilde{g}^2\phi^3(1+z)^2\\ \nonumber &+&2\tilde\kappa\phi^3\phi'^2(1+z)^4   -\big(1-3\tilde\kappa\phi^4(1+z)^4\big)\frac{z'}{1+z}\phi'=0,
\eea
where $\Omega^0_x$ are the present-day density parameters. Given the fact that $z(\tilde t_0)=0$ and $\sum_i\Omega_i(\tilde t_0)=1$ (for a flat universe\footnote{Due to the restrictive observational upper bound on the value of the spatial curvature density parameter $\Omega_{\rm K}<0.005$ \cite{Ade:2015rim}, here, we assume that the background geometry is spatially flat.}),
one can read $\phi'(\tilde t_0)$ from \eqref{frid1} as
\be\label{dphi}
\phi'(\tilde t_0)=\pm\sqrt{\frac{2(1-\Omega_{\rm m}^0-\Omega_{\rm r}^0)-\tilde{g}^2\phi_0^4}{(1+\tilde\kappa\phi^4_0)}}.
\ee
Moreover, to ensure the flatness of the Universe, we require that
\be\label{om-lambda}
\Omega_{G}^0=1-\Omega_{\rm r}^0-\Omega_{\rm m}^0,
\ee
which after fixing the values of the parameters to the observational best values $\Omega_m^0=0.308\pm0.012$ and $\Omega_r^0\sim10^{-4}$, gives $\Omega_{\rm G}^0=0.69$ \cite{Ade:2015rim}. Our gaugessence model has three free parameters $\tilde \kappa$, $\tilde g$ and $\phi_0$ as well as the sign of $\phi'_0$.

From the combination of \eqref{omegaG} and (\ref{frid2}), we obtain
\be\label{ddz}
\frac{z''}{1+z}=3(\frac{z'}{1+z})^2\bigg(1-\frac{\tilde{\kappa}}{9}\big((1+z)^4(\phi^3)_{z}\big)^2\bigg)-\frac12\Omega_{\rm m}^0(1+z)^3,
\ee
in which the subscript ``$z$'' denotes a derivative with respect to $z$. While numerical analysis is necessary to find the exact predictions of the model, yet approximate analytical solutions are possible in some limits. In the following, we integrate the above equations analytically for the matter and dark energy epochs and determine the form of $\phi(z)$. Later, on the next section, we report the result of full numerical study of the model.

\subsection{Matter dominated epoch}

During the matter domination, the scale factor is given as $a(t)\propto \tilde t^{\frac23}$ which implies that $z\propto\tilde t^{-\frac23}$. Inserting $z'(z)\simeq-(\Omega^0_{\rm m})^{\frac12}(1+z)^{\frac52}$ in \eqref{ddz}, we obtain this relation for matter era
$$\frac{\tilde{\kappa}}{9}\big((1+z)^4(\phi^3)_{z}\big)^2\ll1.$$
 Recalling the fact that $\tilde\kappa\gg1$, the above relation implies that during the matter dominated epoch $\phi(z)$ is a slow varying function of $z$, \textit{i.e.} $\phi_{z}(z)\ll1$. As the system evolves to lower redshifts during the matter dominated era, then $\phi_{z}$  gradually increases by time. Note that the value of $\phi_{z}$ is related to the value of $\tilde{\kappa}$ and it decreases by increasing $\tilde{\kappa}$. Since $\phi_{z}\ll1$, the density parameters are approximately
\bea
\Omega_{DR}(z)&\simeq&\frac12\tilde g^2\phi^4(1+z)^4,\quad \\ \nonumber \quad \Omega_{DE}(z)&\simeq &\frac{\tilde\kappa}{\tilde g^2}\phi_{z}^2z'^2(1+z)^2\Omega_{DR}(z).
\eea
As $\phi_{z}(z)$ increases by time evolution, one expect that the ratio of $\frac{\Omega_{DE}}{\Omega_{DR}}$ increases as we are approaching the late times.
After solving the field equation of $\phi$ full numerically in the next section, we show that $\Omega_{DR}(z)\gg\Omega_{DE}(z)$ at high redshifts. Hence the gauge field sector effectively acts like a dark radiation at $z\gtrsim5$ and evolves as $\Omega_{\rm G}(z)\simeq\Omega_{\rm DR}(z)\propto a^{-4}$ (see Fig. \ref{fig-om-w}).

\subsection{Dark energy dominated epoch}
At the late time dark energy dominated era and in the small redshif limit $z\lesssim0.1$, the dark energy density parameter $\Omega_{DE}(z)$ is a very slow varying function of $z$, \textit{i.e.} $\Omega_{DE}(z)\simeq1-\Omega_{\rm m}^0$. Using the fact that $\Omega_{\rm r}^0\ll1$, $z'(\tilde t_0)=-1$, combining equations \eqref{om-lambda} and \eqref{omegaG}, we obtain the following differential equation for $\phi$ during the dark energy era
\be\label{late-dphi}
\big(\phi^3\big)_z\simeq\pm\bigg(\frac{6(1-\Omega_{\rm m}^0)}{\tilde\kappa}\bigg)^{\frac12},
\ee
up to the dominate order. Integrating the above equation, one can then find $\phi(z)$
\be
\phi(z)\simeq\phi_0\bigg(1\pm\frac{6^{\frac12}(1-\Omega_{\rm m}^0)^{\frac12}}{\tilde{\kappa}^{\frac12}\phi_0^3}z\bigg)^{\frac13},
\ee
which implies that in the very late time, $\phi$ smoothly increases (decreases) by the redshift, depends on the sign of $\phi'(\tilde t_0)$. The late time value of redshift derivative of the gauge field is related to $\tilde \kappa$, while larger $\tilde{\kappa}$s lead to smaller $\phi_{z}(z)$. Moreover, the gauge field roaming $\Delta\phi(z)$ is proportional to $\frac{1}{\tilde{\kappa}^{\frac12}\phi_0^2}$, as also confirmed by our full numerical study (see Fig. \ref{fig-psi-z}).

\section{Late time accelerated expansion, numerical analysis}\label{latetime}
Up to now, we show that the gaugessence model can describe the late-time accelerated expansion of the Universe. The background trajectory is completely specified by one initial value $\phi_0$ (as well as the sign of $\phi'(\tilde t_0)$) and two parameters $\tilde g$ and $\tilde \kappa$. Here, considering the observational best values $\Omega_m^0=0.308\pm0.012$ and $\Omega_r^0\sim10^{-4}$ \cite{Ade:2015rim}, we explore the parameter space of our model more thoroughly. In the following, we present the result of numerical study of the field equations \eqref{Frid} for different values of $(\phi_0,\tilde g, \tilde\kappa)$. In this section, we only consider systems with positive $\phi'(\tilde t_0)$ (corresponding to $\phi_{z}(z=0)<0$.). However, regardless of the sign of $\phi'(\tilde t_0)$, the gaugessence model can generate the late time accelerated expansion of the Universe. In the next section, we consider both cases and constrain them with the available observational data.

\begin{figure}
\centering
\includegraphics[width=0.48\textwidth]{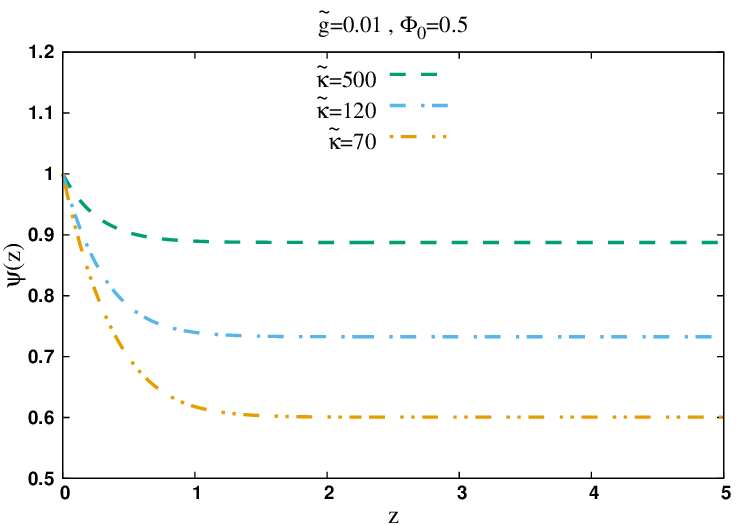}\\ 
\includegraphics[width=0.48\textwidth]{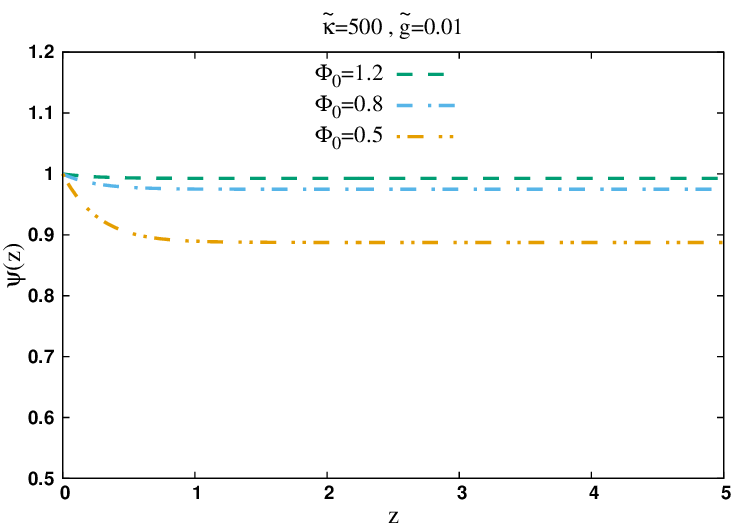}
\caption{The gauge field $\phi(z)$ evolution with respect to redshift, $z$. Here, we normalized each $\phi(z)$ to its own present time value $\phi_0$, $\psi(z)=\frac{\phi(z)}{\phi_0}$. In the upper panel, we set $(\phi_0=0.5, \tilde{g}=0.01)$, while each plot corresponds to a different value of $\tilde{\kappa}$. In the bottom panel, on the other hand, we fixed the value of parameters as $(\tilde{\kappa}=500, \tilde{g}=0.01)$ for systems with three different values of $\phi_0$. 
}\label{fig-psi-z}
\end{figure} 

In the two panels shown in Fig. (\ref{fig-psi-z}), we present the time evolution of the gauge field $\phi$ vs. redshift. The left panel presents $\psi(z)=\frac{\phi(z)}{\phi_0}$ ($\phi_0=\phi(z=0)$) for three different values of $\tilde\kappa$, with $\phi_0=0.5$ and $\tilde g=0.01$. In the right panel, we plot $\psi(z)$ for three different values of the present time field, $\phi_0$, while we fixed the values of the parameters $\tilde\kappa=500$ and $\tilde g=0.01$. These plots confirm our analytical qualitative analysis in the previous section. As we see, $\phi(z)$ is almost constant at high redshifts and during the matter dominated epoch. However, as we approach smaller redshifts, the $\phi'(z)$ increases and thus $\phi(z)$ evaluates faster with the redshift. As we expected from our analytical analysis, the field roaming during its evolution is related to the value of $\tilde\kappa$ and $\phi_0$ and it is proportional to $(\tilde\kappa^{\frac12}\phi_0^2)^{-1}$.

\begin{figure*}
\centering
\includegraphics[width=0.5\textwidth]{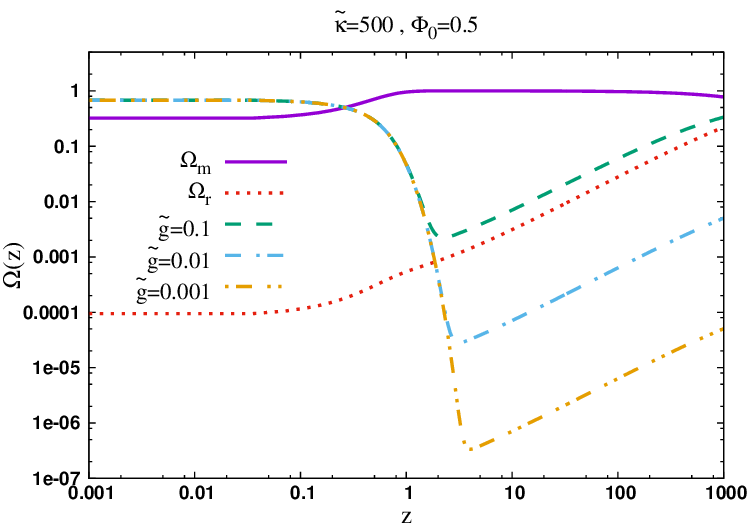}\includegraphics[width=0.5\textwidth]{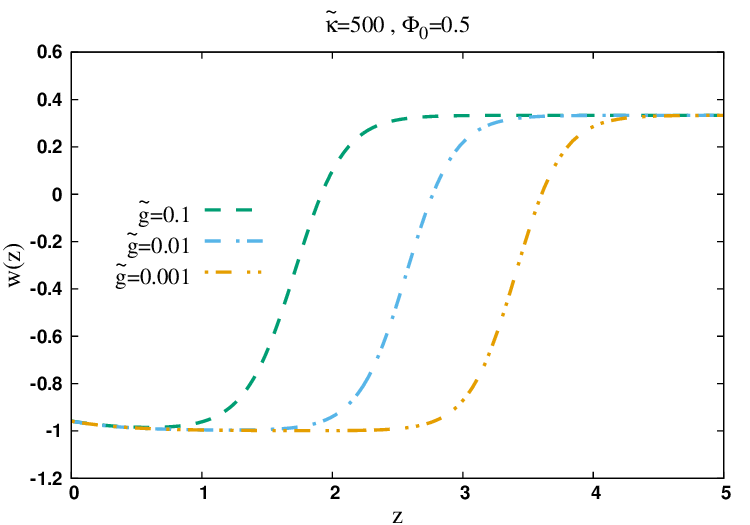}\\
\includegraphics[width=0.5\textwidth]{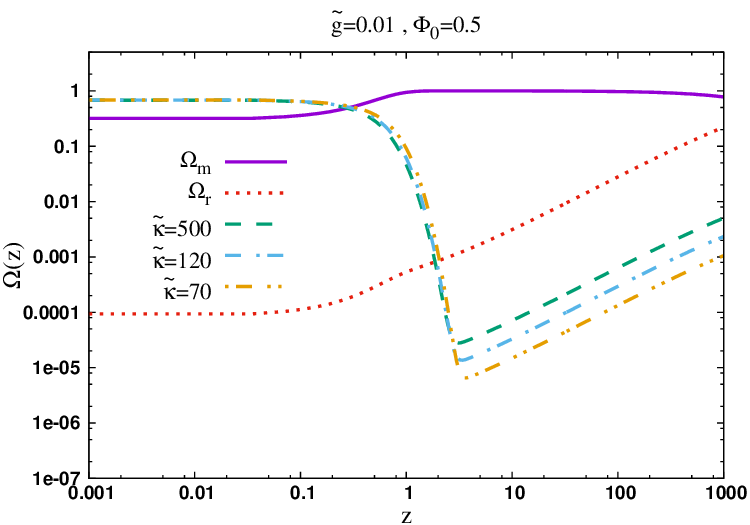}\includegraphics[width=0.5\textwidth]{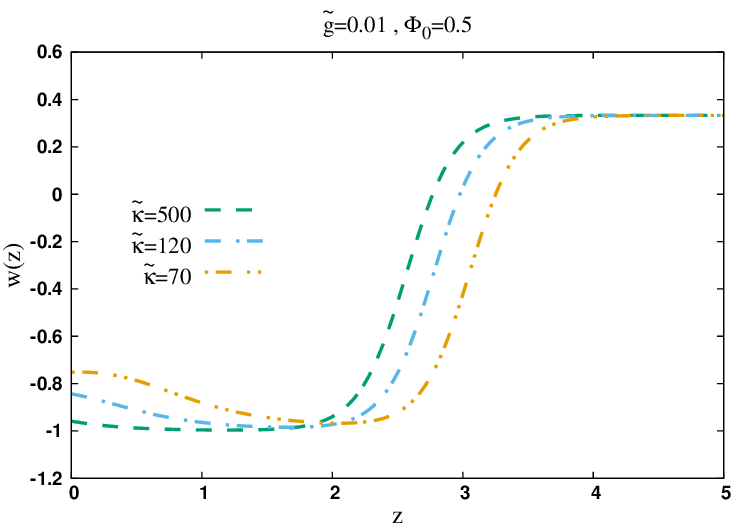}\\
\includegraphics[width=0.5\textwidth]{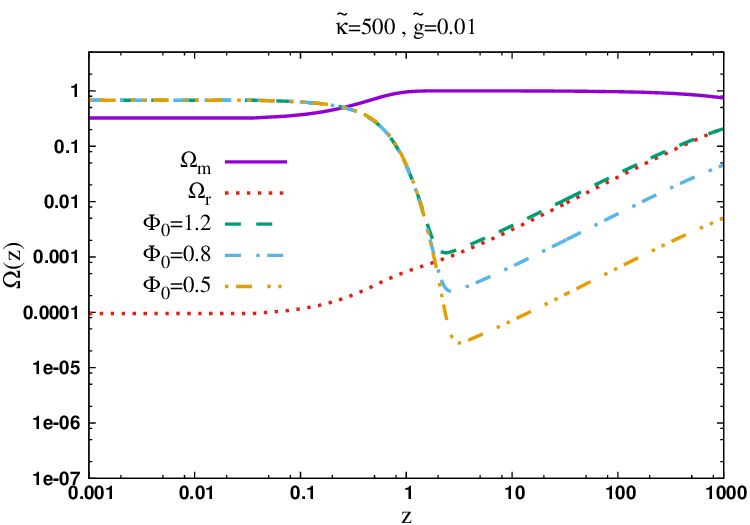}\includegraphics[width=0.5\textwidth]{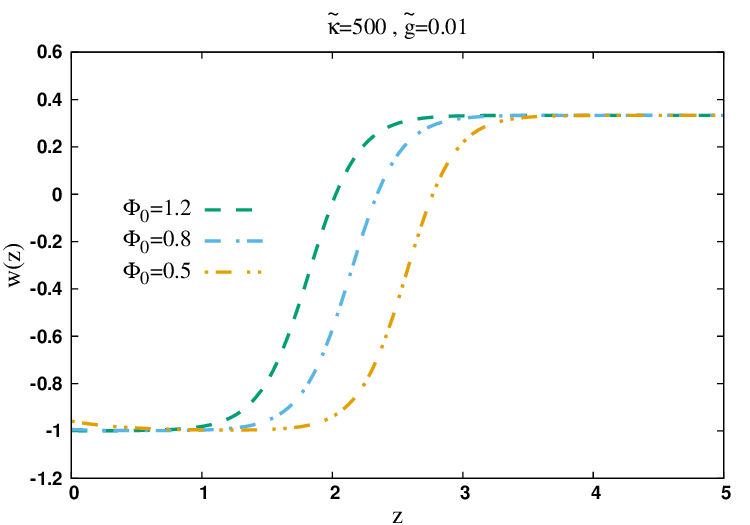}
\caption{Here we explore the  parameter space of the gaugessence model. The left-handed panels show the density parameters for radiation $\Omega_{\rm r}$ (dotted-red line), mater $\Omega_{\rm m}$ (solid-violet line) and $\Omega_{\rm G}$ with respect to redshift $z$. On the right-handed panels we plotted the equation of state of the gauge sector $w(z)$ vs. $z$.}\label{fig-om-w}
\end{figure*} 

We present the evolution of normalized density parameters, (note that in this part we normalize the density parameter by $H^2$ rather than $H_0^2$ see equation (\ref{eq:omega-g})), and EoS of the gauge field, $w$, in the right and left panels of figure (\ref{fig-om-w}) respectively. The \textit{top panels} present systems with different values of $\tilde g$, while the other parameters are fixed ($\tilde\kappa=500$, $\phi_0=0.5$). Although the late time behaviors are roughly the same, the value of $\tilde g$ affects the time evolution at higher redshifts ($z\gg1$). In particular, as we see in the upper-left panel, decreasing $\tilde g$ decreases $\Omega_{G}(z)$ at higher $z$'s which is in form of (dark) radiation $\Omega_{\rm G}\propto a^{-4}$. Note that in all trajectories with $\tilde g<0.01$, the dark radiation density parameter is well below $\Omega_{\rm r}(z)$. The upper-right panel zoomed in on the evolution of EoS in near redshifts. As we see, the gauge field sector is effectively a dark radiation component with EoS $w=\frac13$ at higher redshifts. Then, after a short transition, it takes the form of a dark energy component with EoS $w\simeq-1$. As we see here, the dark energy phase of the gauge field increases by decreasing $\tilde g$.

In the \textit{middle panels}, we fix $\phi_0=0.5$ and ${\tilde g=0.01}$ and investigate how the background trajectories are modified by changing the value of $\tilde\kappa$. The left-middle panel shows that the early $\Omega_{G}(z)$ (at $z\gg1$) evaluate like radiation, $\Omega(z)\propto a^{-4}$, and increases by increasing $\tilde \kappa$. In the middle-right panel we then zoom in on the evolution of EoS in near redshifts. Again, at near redshifts, the gauge field sector is of the form of a dark energy component with EoS $w\simeq-1$, while at higher redshifts it is effectively a dark radiation component with $w=\frac13$. As we see in the right panel, the dark energy phase of the gauge field increases by decreasing $\tilde \kappa$.

Finally, in the \textit{bottom panels} of Fig.(\ref{fig-om-w}), we see systems with different values of $\phi_{0}$, while the other parameters are fixed $\big(\tilde\kappa=500$, $\tilde g^2=0.01\big)$. This plot indicates that increasing $\phi_0$ increases the early time $\Omega_G(z)$ while decreases the time of the dark energy phase of the gauge field sector. The bottom-left panel shows that systems with large field values ($\phi_0\gtrsim1$) make sizeable contribution to the total radiation during the matter era which is not consistent with observational results $\frac{\Omega_{DR}(z)}{\Omega^0_{\rm r}(z)}<0.1$ at high redshifts. (note that with $\big(\tilde\kappa=500$, $\tilde g^2=0.01\big)$ but other area of parameter space may act different). 

\subsection{Generic features of the gaugessence trajectories}

Our numerical analysis shows that our model can generate the late-time accelerated expansion in a large region of the parameter space. Here we summarize the generic features of these solutions in the gaugessence model where $\Omega_{\rm m}^0=0.308\pm0.012$, $\Omega_{\rm G}^0=0.69$ and the critical density $\rho_{\rm c}\equiv3H_0^2\sim \big(10^{-3}\textmd{ev}\big)^4$.
\begin{itemize}

\item Dynamics of the gauge-quentessence trajectories is almost insensitive to its initial field value in early times. In particular, the gauge field component acts like (dark) radiation with EoS of $w=\frac13$ at higher redshifts. After a sharp and fast phase transition around $1<z<5$, it however, takes the form of a dark energy component at late times.

\item There is a range of parameters that the gauge field sector has an energy density a few orders of magnitude smaller than radiation. Due to the cosmic expansion, its energy density then dilutedues like $a^{-4}$ at high redshifts. Thus, the gauge field sector tracks the total radiation energy density during the matter era. As we approach lower redshifts, $\Omega_{G}$ increases and after a quick period of enhancement, it eventually gets constant at the dark energy era. The tracking behavior of the model attracts solutions with different initial values to a common trajectory. That makes the current energy density almost independent of the initial conditions.

\item The gauge field effective value $\phi(z)$ is almost frozen in the matter dominated era and $\phi_{z}$ is negligible. However, as we approach smaller redshifts and around $z\lesssim5$, the value of $\phi'(z)$ increases and thus $\phi(z)$ evaluates faster with the redshift. The field roaming, $\Delta\phi(z)$, during its evolution is related to the value of $\tilde\kappa$ and $\phi_0$ and it is proportional to $(\tilde\kappa^{\frac12}\phi_0^2)^{-1}$.

\item The energy density parameter of the gauge field sector at early times is approximately given as $\Omega_{G}\simeq\frac{\tilde g^2\phi^4_{0}}{a^4}$. Thus, the early $\Omega_{G}(z)$ has a negligible $\tilde\kappa$ dependent. 

\end{itemize}

\section{Contact with observational constraints}\label{data-model}

We now turn to constrain the gaugessence parameters with observational data. In this section, we study the implications of
the supernovae, baryon acoustic oscillation (BAO) and cosmic microwave background (CMB) data sets as well as the current observational upper bounds on the early time dark sectors in our model. Finally, we compare the Taylor expansion of $w$ with observational constraints to further constrain the model.

\subsection{Observational data and best value of parameters}
The observation of distant type-Ia supernovae (SnIa) is one of the most important probes of cosmic expansion. SnIa are one of our standard candles to measure luminosity distance in terms of redshift and shed great light on the late time evolution of the Universe. 
In this part, using SnIa, BAO and CMB data, we find the best value of free parameters of the model $\textbf{P}=\{\Omega_m,H_0,\tilde{\kappa},\tilde{g},\phi_0\}$ as well as their confidence intervals. Here we used the SnIa distance module data from Union 2.1 sample \cite{Suzuki:2011hu} which includes 580 data. The $\chi^2$ is given as
\begin{equation}\label{eq:xi2-sn}
 \chi^2_{\rm sn}=\mathbf{X}_{\rm sn}^{T}\mathbf{C}_{\rm sn}^{-1}\mathbf{X}_{\rm sn}\;,
\end{equation}
where $\mathbf{X}_{\rm sn}=\mu_{\rm th}-\mu_{\rm ob}$ and distance module is 
\be
\mu(z)=5\log_{10}\left[(1+z)\int_0^z\frac{dx}{E(x)}\right]+\mu_0.
\ee
subscripts "ob" and "th" stand for observation and theory respectively. We use covariance matrix $\mathbf{C}_{\rm sn}$ including systematic error from \cite{Union2.1:2012}. Note that in this case our results are marginalized over $\mu_0$.

Baryon acoustic oscillations (BAO) are the frozen leftovers of the oscillations in the relic baryon-photon plasma on the matter power spectrum \cite{Bassett:2009mm}. Since the scale of BAO is very large (with the comoving scale $l_{BAO}\simeq100 \textmd{h}^{-1}\textmd{Mpc}$), up to a very good approximation, it is governed by linear, well-understood physics. That made BAO the next-most robust cosmological probe after CMB fluctuations. In fact, BAO is a powerful standard measure which can provide both angular distance, $D_{A}(z)$,  and Hubble parameter, $H(z)$, using almost linear physics and offers constraints on the background evolution of dark energy. 

\begin{table}
\small{
 \centering
 \begin{tabular}{|c|c|c|}
 \hline
 $z$     & $d(z)$    & ${\rm Survey}$  \\ \hline
 $0.106$ & $0.336$  & {\rm 6dFGS} \citep{Beutler:2011hx}\\ \hline
 $0.35$   & $0.113$ & {\rm SDSS-DR7} \citep{Padmanabhan:2012hf}\\
 $0.57$  & $0.073$ & {\rm SDSS-DR9} \citep{Anderson:2012sa} \\ \hline
 $0.44$  & $0.0916$ &{\rm WiggleZ} \citep{Blake:2011en} \\
 $0.6$   & $0.0726$ & {\rm WiggleZ} \citep{Blake:2011en} \\
 $0.73$  & $0.0592$ & {\rm WiggleZ} \citep{Blake:2011en} \\
 \hline
 \end{tabular}
 \caption{The BAO data which are used in this analyze.}
 \label{tab:bao}
 }
\end{table}

Here, it is useful to introduce the following distance ratio
\begin{equation}\label{eq:d(z)}
 d(z)=\frac{r_{\rm s}(z_{\rm d})}{D_V(z)}\;,
\end{equation}
in which $z_d$ is the redshift of drag epoch, $D_{\rm V}(z)$ is a geometric estimate of the effective distance, defined as 
\begin{equation}\label{eq:dv-bao}
 D_V(z)=\bigg((1+z)^{2}D^{2}_{\rm A}(z)\frac{z}{H(z)}\bigg)^{\frac{1}{3}}\;,
\end{equation}
and $r_{\rm s}(a)$ is the comoving sound horizon
\begin{equation}\label{eq:com-sound-H}
 r_{\rm s}(a)=\int_0^{a}\frac{c_{\rm s}da}{a^2H(a)} \quad \textmd{where} \quad c_{\rm s}(a)=\frac{1}{\sqrt{3(1+\frac{3\Omega_b^0}{4\Omega_{\gamma}^0}a)}},
\end{equation}
where $c_{\rm s}$ is the baryon sound speed. We set the $z_{\rm d}$ fitting formula from \cite{Eisenstein:1997ik}.
BAO has been measured in some different redshits. In Tab.~(\ref{tab:bao}),
we presented the current available data. Following \cite{Hinshaw:2012aka}, we can incorporate the BAO data into $\chi^2$ below
\begin{equation}\label{eq:xi2-bao}
 \chi^2_{\rm bao}=\mathbf{Y}^{T}\mathbf{C}_{\rm bao}^{-1}\mathbf{Y}\;,
\end{equation}
where $\mathbf{Y}$ is given as
\begin{eqnarray}
 \mathbf{Y}&=&(d(0.1)-d_{1},\frac{1}{d(0.35)}-\frac{1}{d_2},\frac{1}{d(0.57)}\\ \nonumber &-&
\frac{1}{d_3},d(0.44)-d_{4},d(0.6)-d_{5},d(0.73)-d_{6}).
\end{eqnarray}
 Moreover we use the covariance matrix $\mathbf{C}_{\rm bao}^{-1}$ which was introduced in \cite{Hinshaw:2012aka}.

For CMB probe we use the so called shift parameter defined by 
\begin{equation}
 R  =  \sqrt{\Omega_{\rm m}^{0}}H_{\rm 0}D_{\rm A}(z_{\ast})\;,
\end{equation}
where $D_{\rm A}$ is the comoving angular diameter distance and $z_{\ast}$ is the red shift at last scattering epoch. 
For $z_{\ast}$ we use the fitting formula from \cite{Hu:1995en} and $\chi^2_{cmb}$ is given by 
\begin{equation}\label{eq:chi2_cmb}
\chi^2_{cmb}=(\frac{R-R_{pl}}{\sigma})^2\;.
\end{equation}  
For $R_{pl}$  and $\sigma$ we use the latest results from {\emph{Planck}} team which is marginalized over the lensing amplitude $(A_L)$ \cite{Ade:2015rim}. The shift parameter provide a simple geometry probe for CMB data but one may think about considering full CMB data to constrain the model which is beyond the scope of this paper.  

Recalling that the overall likelihood function of three individual likelihoods is given by their products, ${\cal L}_{\rm tot}={\cal L}_{\rm sn}\times {\cal L}_{\rm bao}\times {\cal L}_{\rm cmb}$, we obtain the total $\chi^2$ as
\begin{equation}\label{eq:like-tot_chi}
 \chi^2_{\rm tot}=\chi^2_{\rm sn}+\chi^2_{\rm bao} + \chi^2_{\rm cmb}\;.
\end{equation}

\begin{figure*}
\centering
\includegraphics[width=\textwidth]{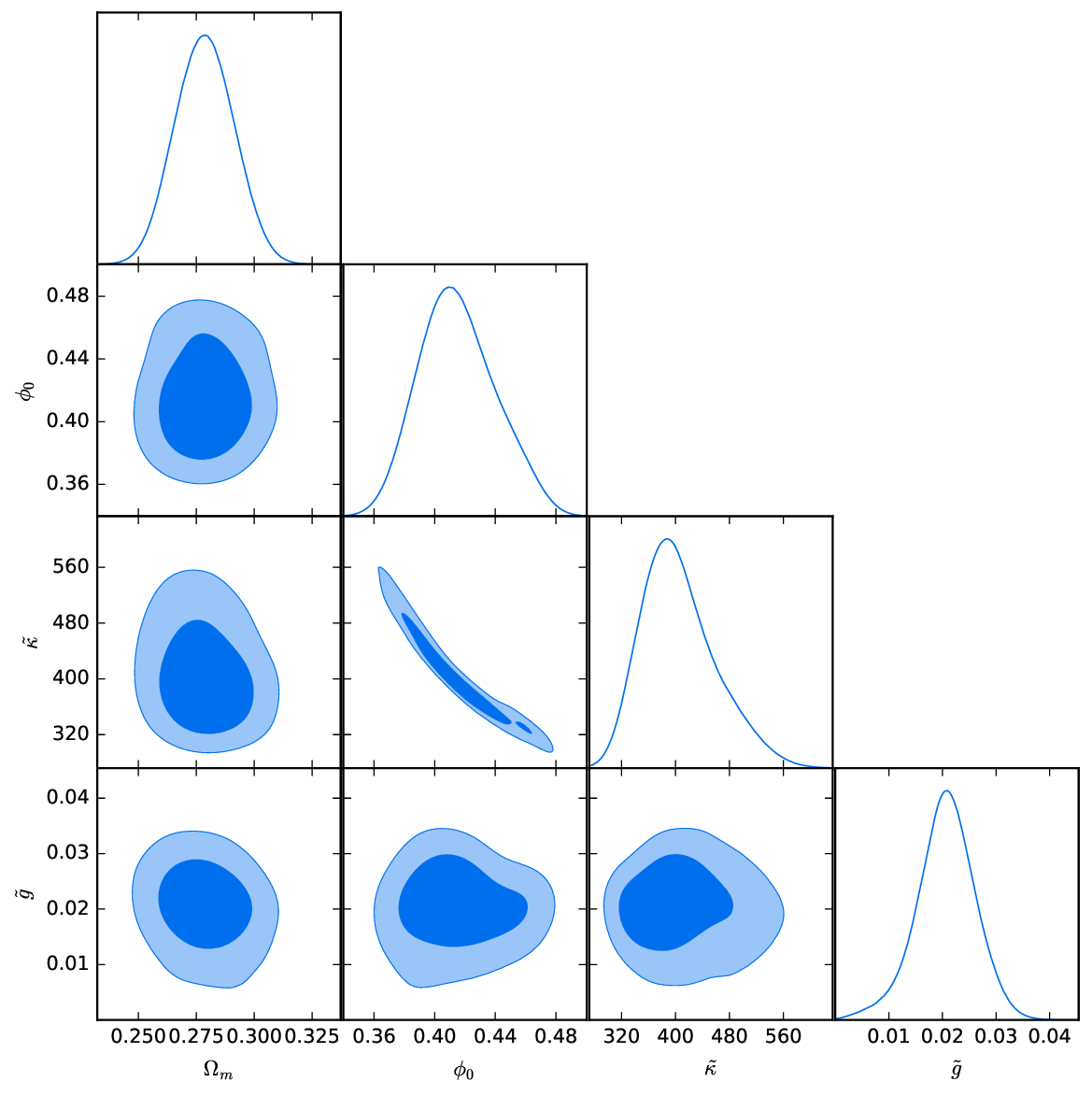}
\caption{1$\sigma$ and 2$\sigma$  confidence regions for the model parameters (model A). The likelihood functions for each parameters are shown by solid line. }\label{fig:cont}
\end{figure*}

We then perform a MCMC (Monte Carlo Markov Chains) analysis to find the minimum of $\chi^2_{\rm tot}$ as well as the best values of parameters and their uncertainties. The results are summarized in Table (\ref{tab:res}) where model A (B) corresponds to systems with positive (negative) values of $\phi'(t_0)$. We present the $1\sigma$ and $2\sigma$ confidence regions and likelihood functions in Fig. (\ref{fig:cont}).  It is noteworthy to mention that although both models are the same in fitting with the data, model B has a relatively bigger $\chi^2_{\rm min}$ comparing to model A. For the sake of comparison we perform our analyzes for the concordance $\Lambda{\rm CDM}$ model and the results is presented in last line of Tab.(\ref{tab:res})

\begin{table*}
\centering
\tabulinesep=2.2mm
\begin{tabular}{|c | c c c  c  c  c| }
\hline 
& $\Omega_m$& $H_0$& $\tilde{\kappa}$ & $\tilde{g}$ & $\phi_0$ & $\chi^2_{min}/nof$ \\ 
\hline
${\rm Model~ A}$ & $0.279\pm 0.013$ &  $70.1\pm 1.4$ &$404^{+40}_{-60}$ & $0.0206^{+0.0055}_{-0.0050}$ & $0.415^{+0.023}_{-0.029}$ & $\frac{544.5}{582}$ \\
\hline
${\rm Model~ B}$ & $0.277\pm 0.016$ &  $68.3\pm 1.9$ &$354^{+120}_{-90}$ & $0.0156^{+0.0065}_{-0.0059}$ & $0.303^{+0.033}_{-0.038}$ & $\frac{559.3}{581}$\\
\hline 
$\Lambda{\rm CDM}$ & $0.295\pm 0.009$ &  $70.5\pm 1.6$ &$-$ & $-$ & $-$ & $\frac{549.1}{585}$\\
\hline
\end{tabular}
\caption{1$\sigma$ limits on our free parameters $\{\Omega_m,H_0,\tilde{\kappa} ,\tilde g, \phi_0\}$ and $\chi^2_{min}$ using the SnIa distance module data from Union 2.1 sample \cite{Suzuki:2011hu}, BAO and CMB data. Model A and B correspond to systems with positive and negative $\phi'(t_0)$ respectively. Results for concordance $\Lambda{\rm CDM}$ is added for sake of comparison.}\label{tab:res}
\end{table*}

\subsection{Dark energy density at early times}\label{early-Omega}
One of the main goals in understanding dark energy is to measure $\Omega_{DE}(z)$ in different redshifts. Up to now, we mainly focus on the late time effects of DE. However, the early time dark energy (EDE), the amount of DE presented in the early times $\Omega_{e}$, can influence CMB anisotropies, structure formation and CMB lensing. In particular, EDE reduces the growth of structures after last scattering and generates less clusters comparing to $\Lambda$CDM. That then leads to a weaker lensing potential to influence the high $l$ anisotropies. The current upper bound on the abundance of DE density in the last scattering surface is $\Omega_{e}(z_{\rm lss})\lesssim0.01$ ($95\%$ CL) \cite{Ade:2015rim}. It is noteworthy to mention that for $\Lambda$CDM, the early contribution of dark energy is $\Omega_{\Lambda}(z_{\rm lss})\approx10^{-9}$.

Any forms of early relativistic dark sectors, dark radiation (DR), is observable through its contribution to the radiation density in the universe. The energy contribution of the early relativistic degrees of freedom is parametrized by the effective number of light species $N_{\rm eff}$, as
\be\label{Neff}
N_{\rm eff}\equiv\frac87\big(\frac{11}{4}\big)^{\frac43}\frac{\rho_{\rm light}}{\rho_{r}}
\ee 
where $\rho_{\rm r}$ is the energy density of photons, while $\rho_{\rm light}$ is the total relativistic energy density in neutrinos and DR at $T\ll 1$ Mev respectively. Since neutrinos are slightly heated and not fully decoupled at the electron-positron annihilation, the cosmological prediction for effective number of nuetrinos is $N_{\nu}=3.046$ \cite{Mangano:2001iu}. Thus, in case that all the relativistic degrees of freedom are SM neutrinos, the numerical factor is $N_{\rm eff}=3.046$. However, observational data constraints the effective relativistic degrees of freedom as $N_{\rm eff}=3.15\pm0.23$ \cite{Ade:2015rim}. Although that number is consistent with the standard value of $3.046$, a significant density of extra radiation is still allowed which may indicates new physics, $\Delta N_{\rm eff}=0.1\pm0.23$. On the other hand, from \eqref{Neff} and after using the SM neutrinos $N_{\nu}$, we have the following relation for the effective (dark) radiation degrees of freedom $\Delta N_{\rm eff}$ 
\be\label{deltaN}
\Delta N_{\rm eff}\simeq\frac32\frac{\Omega_{\rm DR}}{\Omega_{\rm r}}.
\ee
In other words, the Yang-Mills sector of the dark gauge field contributes to the effective relativistic degrees of freedom.
Using the current observational constraint ($\Delta N_{\rm eff}=0.1\pm0.23$), we obtain an upper bound on the ratio of DR density to radiation density at early times, $\frac{\Omega_{\rm DR}}{\Omega_{\rm r}}\lesssim0.2$. 

Coming back to our model, as we showed in the previous section, the gauge field sector acts like a dark energy term with $w\simeq-1$ at late times $z\lesssim5$. However, at the early times $z\gtrsim 5$, it behaves like a (dark) radiation component with EoS equal to $w=\frac13$. Thus, in the gaugessence trajectories, the early DE density $\Omega_{e}=0$, while the gauge field generates an early radiation density $\Omega_{G}=\Omega_{DR}$ at high redshifts. As indicated by our numerical analysis, the gauge field $\phi$ is almost constant at redshifts higher than $10$ ($\phi'(z)\simeq0$ where $z\gg1$), while the total field roaming $\frac{\Delta\phi}{\phi_0}$ (from early times until present) is about 20$\%$. That then leads to the following approximations for the early time energy density parameters
\be
\Omega_{\rm G}(z)\simeq\Omega_{\rm DR}(z)\sim\frac12\frac{\tilde g^2\phi_0^4}{a^4} \quad \textmd{and} \quad \Omega_{\rm r}(z)=\frac{\Omega^0_{\rm r}}{a^4}.
\ee 
Using the observational upper bound on the value of $\Delta N_{\rm eff}$ and $\Omega^0_{\rm r}$, we then obtain 
\be\label{Cnt1}
\tilde g^2\phi_0^4\simeq\frac43\Omega^0_{\rm r}\Delta N_{\rm eff}\lesssim 10^{-5}.
\ee
In the following, we use the above upper bound to further constrain the parameter space.

\subsection{Taylor expansion of $w$}
One of the key points in understanding the nature of the late-time accelerated expansion is the value of $w$. This quantity is equal to $-1$ for the cosmological constant and any deviation from that implies new physics.
As a quintessence model, gaugessence has a redshift dependent equation of state. As we showed in the previous section, $w$ is $\frac13$ at high redshifts and then after a short phase transition, EoS gets close to $-1$. After $z=1$, the EoS is very slow varing and $w(z)$ is well described by its 1st order Taylor expansion around $z=0$. In order to test the time-varying EoS, we expand $w(z)$ in a Taylor series, as discussed in \cite{Ade:2015rim}  
\be\label{Taylor-w}
w(z)\simeq w_0+w_a \frac{z}{1+z}.
\ee
Expanding $w(z)$ in \eqref{w} and after using the fact that $\tilde g^2\phi_0^4\ll\phi'^2_0$, we obtain $\delta w_0$ and $w_a$ as
\bse\label{w0-wa}
\bea
&w_0=-1+\delta w_0\quad \textmd{where} \quad \delta w_0=\frac43\frac{1}{\kappa\phi_0^4},\\
&w_a^{A,B}=\frac{16}{3}\frac{(\frac{\phi'_0}{\phi_0}-1)\kappa\phi_0^4}{(1+\kappa\phi_0^4)^2}\simeq
4(\frac{\sigma_{A,B}\delta w_0^{\frac12}}{\phi_0}-1)\delta w_0,
\eea
\ese
where $\sigma_{A,B}=\pm1$
and in the last equality in the 2nd line, we used \eqref{dphi} and the fact that $\kappa\phi_0^4\gg1$.
Note that in our model, $w_0>-1$ and hence $\delta w_0$ is always positive.

The $1\sigma$ constraints of \emph{Planck}+BAO+SN-Ia+$H_0$ ($68\%$ CL) \cite{Ade:2015rim} gives the current constraints as
$-0.1<\delta w_0<0.2$ and $-1<w_a<0.2$.
Considering the best value of parameters in model A and B we found $(\delta\omega_0= 0.11~,~w_{a}=-0.08)$ and  $(\delta\omega_0=0.44~,~w_{a}=-5.64)$ respectively. These results also confirm that model A is more consistent with observational data than model B.

\section{A quick treatment of the perturbations}\label{pert}

Up to now, we have studied the background evolution of the gaugessence model, including $w(z)$ and $\Omega_{DE}(z)$. Although, our model is close to $\Lambda$CDM at the background level, perturbations may still evolve differently and provides a fingerprint for the model. In particular, the (dark) gauge field sector contributes to the cosmic perturbations and may affect the galaxy clustering and weak lensing. 
In this sector, we investigate the qualitative behavior of the linear cosmic perturbation of our model which capture the main features of the models. The precise full numeric study of the fluctuations is beyond the scope of the present paper and we postpone it for future work. 

As the large-scale structure and weak lensing surveys are measuring the matter-density contrast and the gravitational potentials, we therefore focus on scalar perturbations here. Using the Newtonian gauge, the scalar sector of the perturbed FRW metric can be parametrized as
\be
ds^2=-(1+2\Phi)dt^2+a^2(1-2\Psi)\delta_{ij}dx^idx^j.
\ee
Thus, the space-time metric is fully specified by three quantities $\{H(t),\Psi(t,\vec{x}),\Phi(t,\vec{x})\}$: two Bardeen potentials, $\Psi$ and $\Phi$, as well as the Hubble parameter, given as (neglecting radiation)
\bea\label{Hp}
H^2(z)=&H^2_0&\bigg(\Omega_{\rm m}^0(1+z)^{3}+(1-\Omega_{\rm m}^0)\\ \nonumber &\exp &\big(3\int^z_{0}(1+w(z'))d\ln(1+z)\big)\bigg),
\eea
where $w$ is the EoS of dark sector. Note that $\Phi$ is the Newtonian potential and in the absence of anisotropic stress in general relativity, e.g. $\Lambda$CDM, we have $\Psi=\Phi$. However, in most of modified gravity models and some dark energy models including vector and gauge fields, we may have a non zero anisotropic stress.

\subsection{Cosmic structure formation and the dark sector}

Cosmic structure formation is another way to explore the dark energy models besides the expansion history. Linear perturbations governed by the well-known physics and free of many astrophysical complexities. We can write the small fluctuations to the homogeneous energy density ($\bar{\rho}_{m}(t)$) and the Hubble flow velocity $Hx^i$ as below
\be
\rho_{m}(t,\vec{x})=(1+\delta_{m}(t,\vec{x}))\bar{\rho}_{m}(t), ~ v^i=Hx^i+\partial_i u(t,\vec{x}),
\ee
where overbars denotes background quantities,$\delta_m$ is the fractional density contrast of the dark matter and $\vec{u}$ is its peculiar velocity. The linear order continuity and Euler equations govern the evolution of $\delta_m$ and $u_m$ which are respectively as below
\be\label{linear-eq}
\dot{\delta}_{m}+\frac{1}{a}\partial^2u_{m}=0 \quad \textmd{and} \quad \dot u_{m}+H u_{m}+\frac1a\Psi=0.
\ee
Moreover, the Poisson equation gives $\Psi$ in terms of the total energy density as 
\bea\label{poisson}
k^2\Psi=-4\pi Ga^2\bar\rho_{\rm m}\bigg(\delta_{m}+\frac{\bar\rho_{\rm G}}{\bar\rho_{\rm m}}\delta_{G}\bigg),
\eea
where $\delta_{G}$ is the density contrast of the dark gauge field \cite{Amendola:2012ys}. From the combination of \eqref{linear-eq} and \eqref{poisson}, we then obtain a dynamical field equation for $\delta_{m}$
\bea\label{delta-m}
&\ddot{\delta}_{m}&+2H\dot{\delta}_m=4\pi G\bar\rho_{\rm m}Q(t)\delta_{m},
\eea
where $Q-1$ is the deviation of dark matter evolution from $\Lambda$CDM prediction. In our model, around $z\lesssim5$ the gauge field sector effectively acts like a dark energy sector with $w_{G}\simeq-1$. Moreover, 
the ratio $\frac{\bar\rho_{\rm G}}{\bar\rho_{\rm m}}$ is negligible for $z\gtrsim1$ which implies that $Q(z)-1$ is almost zero at that regime and 
the Hubble parameter \eqref{Hp} is equal to the $\Lambda$CDM.
  Thus, during the matter era, the structure formation is almost exactly the same as $\Lambda$CDM, $\delta_m\propto a$ in our model. On the other hand, around $z=0.5$, dark energy dominates over matter and may change the rate of structure formation.
 Similar to the standard quintessence models, our model has a sound speed very close to one, and does not cluster significantly inside the horizon. Therefore, our dark energy is very smooth comparing with the cold dark matter, so $\delta_G$ is negligible.

Although our dark energy does not cluster, however, it can still affect the dark matter clustering by changing the Hubble parameter comparing with the $\Lambda$CDM \cite{Mehrabi:2015kta,Mehrabi:2015hva}. Investigation of growth of structures in our model needs a careful study which we postpone to the future work. 
 The other way that our dark energy model can affect the clustering is by generating a non-zero anisotropic stress which we discuss in the following.

\subsection{Anisotropic stress}

The gravitational lensing is sensitive to the combination of $\Phi+\Psi$, while the anisotropic clustering of galaxies in redshift space measures the peculiar velocity which is related to $\Psi$. Thus, the combination of them makes us able to determine both $\Psi$ and $\Phi$. From the theory on the other hand, the off-diagonal spatial part of the linear perturbed Einstein equation gives
\be
\Psi=\Phi+\Pi^s,
\ee
where $\Pi^s$ is the scalar anisotropic stress, given as $\delta T^i_{j}-\frac13\delta^i_{j}\delta T^i_j=(\partial_{ij}-\frac13\bigtriangledown^2)\Pi^s$. As a result, presence of anisotropic stress leads to inequality of the Bardeen potentials. Non-vanishing anisotropic stress is usually considered as a smoking gun for modified gravity, however,\textit{ in principle}, it can be generated by higher spin dark energy models, e.g. dark vector and gauge fields\footnote{During the radiation era, the relativistic particles, i.e. photons, neutrinos and dark radiation contributes to $\Pi^s$, generating a non-zero scalar anisotropic stress. The background cosmic neutrino fluctuations and any possible free-streaming particles generate significant anisotropic stress which induces a characteristic phase shift in the acoustic peaks of CMB \cite{Baumann:2015rya}. This phase shift recently has been detected in \cite{Follin:2015hya} and future CMB observations will improve our understanding about them.}. In our model, the dark sector during early matter era ($z\gtrsim10$) acts like dark radiation, decays like $a^{-4}$ and is negligible. At $z\lesssim5$, on the other hand, it acts like a smooth dark energy component without anisotropic stress. As a result, the dark gauge field does not generate anisotropic stress during the matter and dark energy eras, we have $\Psi=\Phi$.

\section{Discussion and Conclusions}\label{conclusion}

In this paper, we investigated the cosmological evolution of a novel quintessence scenario, gaugessence. Our dark energy candidate is a self-interacting (dark) gauge field which interacts gravitationally with the visible Universe and minimally coupled to Einstein gravity. The gauge field theory of the model consists of the standard Yang-Mills with an EoS equal to $\frac13$, and $\kappa(F\tilde F)^2$ term which effectively is a dark energy sector. The (dark) gauge sector has two free parameters, the gauge field coupling $g$ and the dimensionful $\kappa$ which has the dimension of $\textmd{mass}^{-4}$. Thus, in the regime that the $\kappa$-term dominates, the dark gauge sector acts like a quintessence model, justifying the name gaugessence. Our gauge field theory, originally, has been introduced and studied as an inflationary model to describe the early time exponential expansion after the big bang in \cite{Maleknejad:2011jw, Maleknejad:2011sq}, as gauge-flation.
Here, we examined the late time cosmology of the model in the presence of matter and radiation to explain the late time accelerated expansion of the Universe.  As discussed in detail in \cite{Maleknejad:2011sq} and \cite{SheikhJabbari:2012qf}, the $(F\wedge F)^2$ term can be generated as an effective field theory by integrating out the massive axion field ($m_{\chi}\sim H_0$) in a theory with a Chern-Simons interaction between the gauge field and the axion. The value of the axion mass can be found by $
	\frac12m^2\chi_0^2\sim3M_{\rm pl}^2H_0^2$, which gives $m_{\chi}\sim H_0$. The $\kappa$ parameter and the cut-off scale of the effective theory $\Lambda$ are related as $\kappa\sim \Lambda^{-4}$. Recalling that (after recovering $M_{\rm pl}$) $\kappa=\frac{\tilde{\kappa}}{\tilde g^2}\big(\frac{M_{\rm pl}}{H_0}\big)^2\frac{1}{M_{\rm Pl}^4}$, our best fit values gives a cut-off value equals to $\Lambda\lesssim 1\rm meV=10^{30}~H_0$. As a result, our effective theory works in $H\lesssim 0.1~\Lambda$ which corresponds to redshifts $z\lesssim 10^{16}$ and $T\lesssim 100~\rm Gev$. Therefore, our effective theory is reliable around and after EW phase transition until the present time.

We performed a likelihood analysis to compare the available SnIa,  BAO and CMB data with the gaugessence model and find the best values of our parameters. During the radiation and early matter eras, the gauge field sector acts like (dark) radiation and dilutes like $a^{-4}$. We then have a quick transition period in which $w$ sharply decreases from $\frac13$ to $-1$, while its energy density parameter increases rapidly. Eventually, at low-redshifts, the gauge field sector takes the form of a dark energy and its energy density becomes the dominant part of the energy budget of the Universe, $\Omega_{G}\simeq0.7$. Due to the tracking feature of our model, solutions with different initial values are attracted to a common trajectory. That makes the current energy density almost independent of the initial conditions. The effective field value $\phi(z)$ is almost frozen during the radiation and matter eras, while as we approach smaller redshifts ($z\lesssim5$), it starts to evaluate faster with time. The total field roaming of the gauge field, $\Delta\phi(z)$, is about or less than 10$\%$ for the acceptable systems.

The presence of early dark radiation (EDR) is a robust prediction of our model. Our gauge quintessence model act like a radiation component at early time and may have a contribution to the total radiation energy density. During the recombination era, the energy density of EDR  manifests itself observationally as a contribution to the effective number of neutrino species, $N_{\rm eff}$. The current constraint on the effective relativistic degrees of freedom from the Planck satellite is $N_{\rm eff}=3.15\pm0.23$ \cite{Ade:2015xua}, while the SM perdition is $N^{SM}_{\rm eff}=3.046$. That then leads to $\Delta N_{\rm eff}=0.1\pm0.23$ which may indicates new relativistic degrees of freedom at the time of recombination. Future CMB polarization experiments will improve our constraints on $\Delta N_{\rm eff}$ by one or two order of magnitude \cite{Wu:2014hta}. Using the current observational constraint on $\Delta N_{\rm eff}$ to our model, we obtain $g^2\phi^4_0\lesssim10^{-5}H_0^2$.


\acknowledgments
We thank the anonymous referee whose comments helped to improve the paper. In addition,
we greatly appreciate M. M. Sheikh-Jabbari for his valuable comments on the manuscript.
Work of A. MN. is supported in part by the grant from \textit{Boniad Melli Nokhbegan of Iran}.  
A. MN. acknowledges the hospitality of the Aspen Center for Physics and support from PHY-1066293.

\bibliographystyle{apsrev4-1} 
\bibliography{ref}
  
 \end{document}